\documentclass[%
preprint,
 amsmath,
 amssymb,
 aps,
 prl,
]{revtex4-2}
\usepackage[colorlinks=true,linkcolor=red,citecolor=blue]{hyperref}
\usepackage{graphicx}
\usepackage{dcolumn}
\usepackage{bm}

\usepackage[normalem]{ulem}
\newcommand{\stkout}[1]{\ifmmode\text{\sout{\ensuremath{#1}}}\else\sout{#1}\fi}

\begin{document}

\title{Intermittency Signatures in the Deformation of a Passive Droplet in Active Turbulence}

\author{Sudeep Halder}
\email{ph21065@iisermohali.ac.in}
\affiliation{Indian Institute of Science Education and Research Mohali, Knowledge City,
Sector 81, SAS Nagar – 140306, Punjab, India}
\author{Abhishek Chaudhuri}
\email{abhishek@iisermohali.ac.in}
\affiliation{Indian Institute of Science Education and Research Mohali, Knowledge City,
Sector 81, SAS Nagar – 140306, Punjab, India}

\begin{abstract}
We use fully resolved nematohydrodynamic simulations to study deformation statistics of a passive nematic droplet in two-dimensional extensile active-nematic turbulence. We find that the droplet aspect ratio serves as a scalar probe of the active bath. Its increments show heavy-tailed distributions with dependence on the time lag, scale-free burst statistics and multiscaling structure functions which establish temporal intermittency. While the mean deformation increases with activity, normalized intermittency is strongest at lower activity. This suggests slower and more coherent bath forcing. When compared with translational and forcing-side fluctuations, it reveals a hierarchy of intermittency: shape is more weakly intermittent than translation and active-stress fluctuations, consistent with filtering by interfacial restoring forces. Power spectra show an extended near-$1/\omega$ regime for the maximal normal interface velocity, distinct from the steeper, approximately $1/\omega^{2}$ spectrum of the interfacial active stress. Soft inclusions thus reveal how interfacial restoring forces convert active forcing into bursty, scale-rich deformation dynamics.
\end{abstract}

\maketitle

Active fluids are driven by a continuous injection of energy at the microscopic scale, and they show chaotic spatiotemporal flows known as active turbulence~\cite{ramaswamy2010mechanics,marchetti2013hydrodynamics,thampi2016active,alert2022active,guasto2012fluid,saintillan2018rheology}. This is observed in various overdamped systems, from bacterial suspensions and dense swimmer collectives to microtubule--motor mixtures and epithelial tissues~\cite{cisneros2007fluid,sokolov2007concentration,schaller2010polar,ishikawa2011energy,creppy2015turbulence,wensink2012meso,guillamat2017taming,martinez2021scaling,doostmohammadi2018active,doostmohammadi2017onset}. While active turbulence is compared phenomenologically to inertial turbulence, their origin, scaling structure and defect dynamics are fundamentally different~\cite{alert2020universal,thampi2014instabilities,giomi2013defect,thampi2013velocity,giomi2015geometry,doostmohammadi2019coherent,kozhukhov2022mesoscopic,gautam2024harnessing,guillamat2018active,young2021many,ruske2021morphology,zhao2024asymmetric,hernandez2021poisson}.  Active turbulence can be divided into various classes depending on their orientational order (polar or nematic) and if momentum is conserved (wet or dry), and these classes differ in their route to chaos and their scaling properties~\cite{alert2022active}. In this study, we look at a two-dimensional extensile wet active nematic, where the spatiotemporally chaotic flow is organized by motile disclinations~\cite{thampi2016active}.

A central question is whether the fluctuations generated by active turbulence are effectively Gaussian or show intermittency, i.e., rare, intense events that dominate higher-order statistics. In inertial turbulence, intermittency is observed from the heavy-tailed distributions of increments and the anomalous structure-function scaling~\cite{frisch1995turbulence}. Such features are stronger in the statistics of Lagrangian tracer particles~\cite{yeung2002lagrangian}. Bursty dynamics and scale-free temporal organizations seen in these systems are also observed in other driven nonequilibrium systems~\cite{bak1988self}. Indeed, intermittency and multiscaling have also been reported in other classes of active turbulence~\cite{mukherjee2023intermittency, kiran2025onset}. While different from the active nematic studied here, these studies establish that scale-free, intermittent statistics are a recurring feature of active flows.

In biological and synthetic systems, active fluids continuously interact with soft boundaries, such as membranes, vesicles, and droplets, which deform under active stresses. While rigid tracers~\cite{wu2000particle,morozov2014enhanced,patteson2016particle,burkholder2017tracer,ray2023rectified,foffano2012colloids} and the phase behavior of active droplets and emulsions~\cite{padhan2023activity,tiribocchi2016shear} have been well studied, the dynamics of \emph{deformable} passive inclusions in active turbulence remain poorly understood. Unlike rigid tracers, soft droplets possess internal shape modes and interfacial restoring forces~\cite{hernandez2021poisson}. This allows them to store, filter, and redistribute active forcing over time~\cite{chandler2025active,hadjifrangiskou2023active,singh2024anomalous,gao2017self,zhao2024asymmetric}.

In earlier work, we showed that a passive nematic droplet in active turbulence undergoes anomalous center-of-mass dynamics and shows activity-dependent deformation amplitudes~\cite{singh2024anomalous}. However, the statistical nature of the deformation dynamics remained unresolved. More precisely, whether the shape fluctuations arise from weakly correlated Gaussian forcing or whether they show intermittency.

In this paper, we use fully resolved nematohydrodynamic simulations of a passive nematic droplet in two-dimensional extensile active turbulence to show that the aspect ratio of the droplet serves as an extended scalar probe of the active bath. Its increments show heavy-tailed, lag-dependent distributions, the structure functions display multiscaling, and deformation bursts obey scale-free statistics. We further find that although the mean deformation increases with activity, intermittency in the normalized fluctuations is strongest at lower activity. This is consistent with slower, more coherent forcing. When we compare deformation fluctuations with translational motion and the active-stress fluctuations at the interface, we find that deformation is more strongly filtered than translation. The observed $1/\omega$ spectrum of the maximal normal interface velocity is also not simply due to the forcing stress of the bath. Our results establish soft inclusions as sensitive probes of intermittency in active turbulence and reveal how interfacial restoring forces transform active forcing into bursty deformation dynamics.

\begin{figure*}[!ht]
	\includegraphics[width=1.0\linewidth]{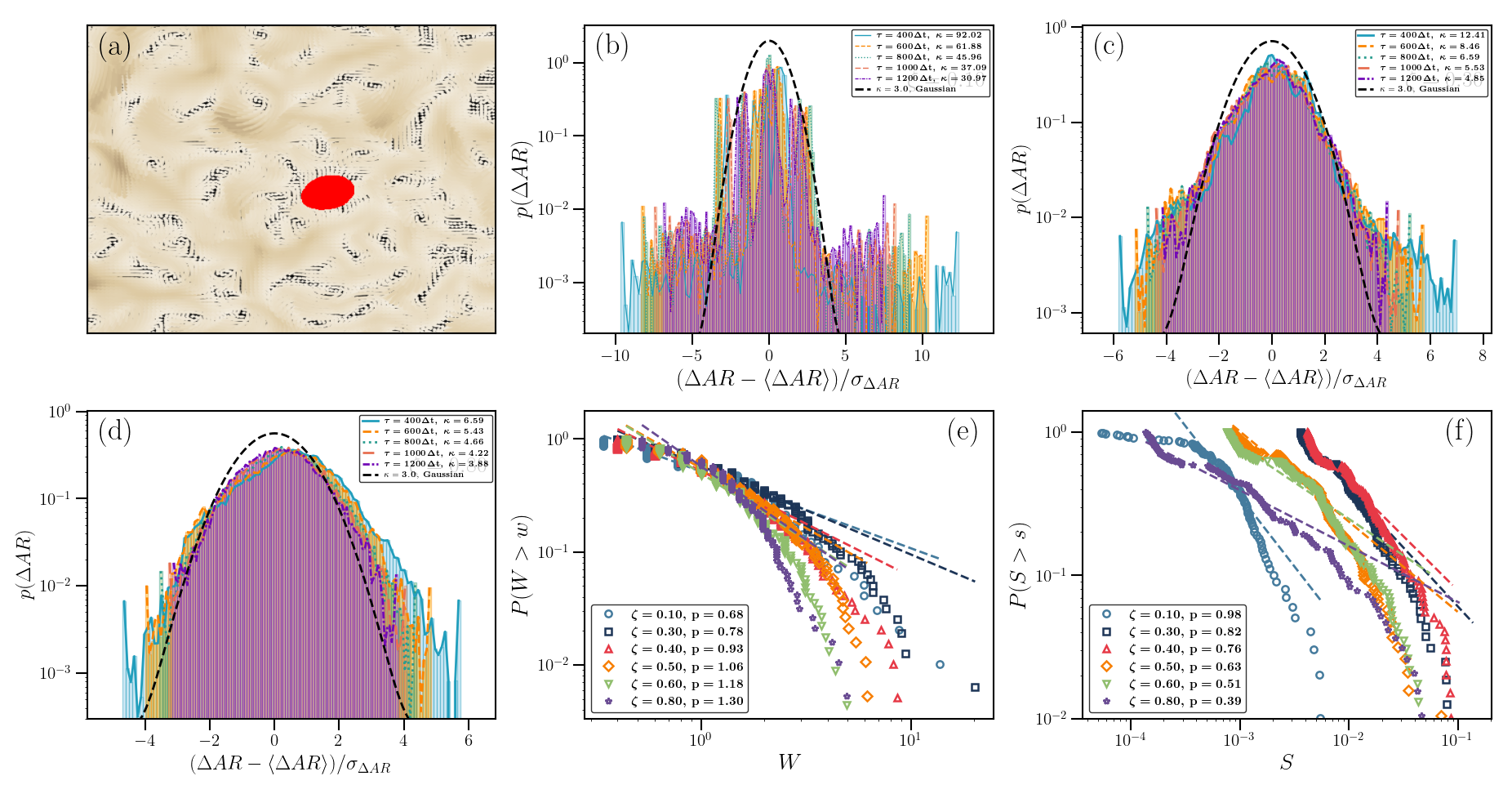}
\caption{Intermittent deformation of a passive droplet. (a) Snapshot of the
deformable droplet (red) in the active turbulent velocity field (cream) at
$t=600$. (b)--(d) Probability density functions $p(\Delta AR)$ of normalised
aspect-ratio increments $(\Delta AR-\langle\Delta AR\rangle)/\sigma_{\Delta AR}$
for $\zeta = 0.10$, $0.30$, and $0.80$, at several lag times $\tau$ (legends);
solid lines are kernel estimates, bars are histograms, legends list the
kurtosis $\kappa$, and the dashed black lines are unit-variance Gaussian ($\kappa = 3$). (e),(f) CCDF of waiting times $W$ and burst sizes $S$ for $\tau = 600\,\Delta t$ and several
activities $\zeta$; dashed lines are power-law fits $P\sim W^{-p_W}$ and
$P\sim S^{-p_S}$, with $p_W$, $p_S$ in the legends (fitting protocol in the SM).}
	\label{fig:AR_increments_bursts}
\end{figure*}

\vskip 0.2cm

\noindent

\emph{Model.} We consider a two-dimensional incompressible active nematic fluid containing a passive nematic droplet~\cite{singh2024anomalous}. The nematic bath is described by Beris-Edwards $Q$-tensor hydrodynamics~\cite{ramaswamy2010mechanics,marchetti2013hydrodynamics,doostmohammadi2018active}, where $Q$ is the symmetric traceless tensor $Q_{\alpha\beta}(\mathbf{x},t)$ characterizing orientational order and the velocity field $\mathbf{u}(\mathbf{x},t)$ satisfies $\nabla\,\cdot\,\mathbf{u}=0$. The evolution of $Q$ is given as
\begin{equation}
\partial_t Q_{\alpha\beta} + u_\gamma \partial_\gamma Q_{\alpha\beta} - S_{\alpha\beta} = \Gamma H_{\alpha\beta},
\label{eq:Qeq_main}
\end{equation}
where $\Gamma$ is the rotational diffusivity, $H_{\alpha\beta}$ the molecular field, and $S_{\alpha\beta}$ the flow-alignment term. We use the one-elastic-constant approximation and a standard quartic free energy, details of which are given in the Supplementary Material (SM). The momentum equation is
\begin{equation}
\rho (\partial_t \mathbf{u} + \mathbf{u}\,\cdot\,\nabla \mathbf{u}) = -\nabla p + \eta \nabla^2 \mathbf{u} + \nabla\,\cdot(\boldsymbol{\sigma}^{\mathrm{P}} + \boldsymbol{\sigma}^{\mathrm{A}}) - \mu \mathbf{u},
\label{eq:NS_main}
\end{equation}
where $\rho$, $\eta$, and $\mu$ are the density, viscosity, and substrate friction. The passive stress $\boldsymbol{\sigma}^{\mathrm{P}}$ is given by the standard Beris--Edwards form~\cite{beris1994thermodynamics}, while the active stress is $\sigma^{\mathrm{A}}_{\alpha\beta}=-\zeta Q_{\alpha\beta}$ with extensile activity $\zeta>0$. We choose parameters such that the droplet-free bath lies deep in the mesoscale turbulent regime~\cite{thampi2016active,doostmohammadi2017onset,singh2024anomalous}.

\begin{figure*}[!ht]
	\centering
	\includegraphics[width=1.0\linewidth]{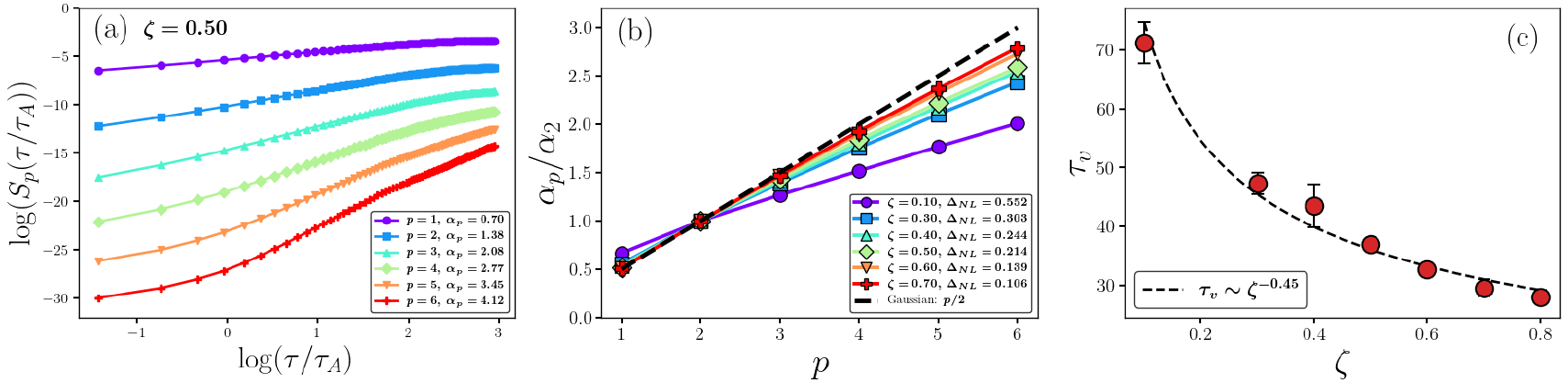}
    \caption{(Color online) Structure functions and multiscaling of droplet-shape increments. (a) Log--log plots of the $p$th-order structure functions $S_p(\tau) = \langle |\Delta AR(t,\tau)|^p \rangle$ versus lag time $\tau/\tau_A$ for a representative activity ($\zeta = 0.50$) and orders $p = 1,\dots,6$ (legend); straight lines are power-law fits $S_p(\tau)\sim\tau^{\alpha_p}$ over an intermediate scaling range, with the fitted $\alpha_p$ listed. (b) Exponents $\alpha_p/\alpha_2$ versus order $p$ for several activities $\zeta$ (legend); the dashed line is the Gaussian self-similar prediction $p/2$. (c) Bath coherence (vortex-lifetime) time $\tau_v$ versus activity, with the fit $\tau_v \sim \zeta^{-0.45}$.}
	\label{fig:structure_functions}
\end{figure*}

We model the passive droplet as a circular nematic region of radius $R$ with zero activity ($\zeta=0$), and hydrodynamically coupled to the surrounding fluid. The interface has line tension $\gamma$ and is represented by a contour $\mathbf{x}_{\mathrm{s}}(s,t)$ advected by the flow, $\partial_t \mathbf{x}_{\mathrm{s}}=\mathbf{u}(\mathbf{x}_{\mathrm{s}},t)$. The interfacial force density is $\mathbf{f}_{\mathrm{I}}(\mathbf{x},t) = \int \gamma \kappa_{\mathrm{n}}(s,t)\, \mathbf{n}(s,t)\,\delta(\mathbf{x}-\mathbf{x}_{\mathrm{s}}(s,t))\, \mathrm{d}s$, where $\kappa_{\mathrm{n}}$ and $\mathbf{n}$ are the local curvature and outward normal~\cite{peskin2002immersed}. This force produces the Laplace pressure jump and enters Eq.~\eqref{eq:NS_main} through $\nabla\,\cdot\boldsymbol{\sigma}^{\mathrm{P}} \to \nabla\,\cdot\boldsymbol{\sigma}^{\mathrm{P}}+\mathbf{f}_{\mathrm{I}}$. The implementation details are provided in the EM.

We nondimensionalise Eqs.~\eqref{eq:Qeq_main}--\eqref{eq:NS_main} using the scales $x_0$ and $t_0$ defined in the EM. The control parameters are dimensionless activity $\zeta$, line-tension number $\gamma t_0/(\eta x_0)$, and droplet radius $R/x_0$. We choose $R$ comparable to the active-vortex size. To characterize the bath, we define the turnover rate $\omega_{\mathrm{A}}=u_{\mathrm{rms}}/\ell_{\mathrm{c}}$ and turnover time $\tau_{\mathrm{A}}=\omega_{\mathrm{A}}^{-1}$, where $u_{\mathrm{rms}}$ and $\ell_{\mathrm{c}}$ are the rms velocity of the fluid and velocity-correlation length. The coupled equations are solved using a finite-volume pressure-projection scheme within an Eulerian one-fluid framework~\cite{unverdi1992front,tryggvason2011direct}. The interface is discretized by $N_{\mathrm{f}}$ Lagrangian points and remeshed periodically. The active bath is first evolved to statistical steady turbulence, after which the droplet is inserted. Statistics are collected over several hundred turnover times.

To quantify deformation, we define the aspect ratio $\mathrm{AR}(t)=a(t)/b(t)$, where $a$ and $b$ are the major and minor axes of the best-fit ellipse. As shown in the SM, the $m=2$ mode contributes more than $90\%$ of the shape variance. Thus $\mathrm{AR}(t)$ provides a compact deformation measure. Intermittency is analysed using increments over lag times $\tau$: $\Delta \mathrm{AR}(\tau;t) = \mathrm{AR}(t+\tau) - \mathrm{AR}(t)$, from which we compute probability distributions, moments, and burst statistics. We also monitor the droplet center of mass, $\mathbf{R}_{\rm cm}(t) = (1/A(t))\int_{\Omega_d(t)}\mathbf{x}\,\mathrm{d}^2x$, the interfacial active stress, $\sigma_{\mathrm{act}}(\mathbf{x},t) = {-}\zeta\, Q_{\alpha\beta}\, n_\alpha n_\beta$, and the center-of-mass speed $U(t) = \left|\frac{\mathrm{d}\mathbf{R}_{\rm cm}}{\mathrm{d}t}\right|$. These quantities allow us to compare deformation intermittency and translational or forcing-side fluctuations.

\vskip 0.2cm
\noindent
\emph{Results.} In this study, we investigate the statistics of droplet deformation. The active bath is a dynamic mix of vortices, jets, and motile $+1/2$ defects. We fix the radius of the droplet while varying the activity $\zeta$. This is to determine whether shape fluctuations remain Gaussian or show temporal intermittency. At all activities, the aspect ratio $\mathrm{AR}(t)$ fluctuates around a mean slightly above unity, with larger fluctuations at higher activity. We analyze increment statistics, bursts, structure functions, and spectra. We find that droplet deformation is strongly intermittent, with normalized intermittency being strongest at lower activity.

\vskip 0.2cm
\noindent
\emph{Increment statistics and deformation bursts.} We look at increments of aspect ratios over lag times $\tau$ over two decades. For every $\tau$ and $\zeta$, we calculate the probability distribution functions (PDFs) of the normalized increments $(\Delta \mathrm{AR}-\langle \Delta \mathrm{AR}\rangle)/\sigma_{\Delta \mathrm{AR}}$. $\sigma_{\Delta \mathrm{AR}}(\tau)$ gives the corresponding standard deviation. In Fig.~\ref{fig:AR_increments_bursts}(b,c,d), we show representative PDFs for low, intermediate, and high activity, respectively. For all the activities, the PDFs are non-Gaussian. They have nearly symmetric cores and pronounced heavy tails. This behavior persists across all $\tau$ values. However, extreme events become less prominent at larger $\tau$. Kurtosis is well above the Gaussian value of $3$. Importantly, while the absolute deformation amplitude increases with activity, the normalized increment shows most intermittency at low $\zeta$ values. This suggests that long quiet intervals are separated by rare large excursions.

To explain this, we define deformation bursts as contiguous intervals where $|\Delta \mathrm{AR}(\tau_b;t)|$ exceeds a threshold value $\theta$, using a base lag $\tau_b = 600\,\Delta t$. This threshold is fixed across activities and is chosen from the low-activity distribution (see SM). For a given burst episode $i$, we define the waiting time $W_i \equiv t_{i+1}^{\rm start}-t_{i}^{\rm end}$ and burst size $S_i \equiv \sum_{t\in {\rm episode}\ i} |\Delta \mathrm{AR}(\tau_b;t)|$. In Fig.~\ref{fig:AR_increments_bursts}(e,f), we plot their complementary cumulative distribution function (CCDF). Over one to two decades, the CCDFs show power law forms: $P(W > w)\sim W^{-p_W}$ and $P(S > s)\sim S^{-p_S}$.

The waiting-time exponent increases with activity. This indicates that an extended period of inactivity between deformation events becomes less likely as the bath becomes more strongly driven. This is consistent with the increment PDFs. At low activity, deformation events are rarer but more extreme, relative to the typical scale of fluctuations. We interpret the burst-size distribution $S_i$ separately since the threshold $\theta$ is held fixed across activities. Note that both the absolute deformation amplitude and the frequency of crossing the threshold increase with $\zeta$. Now, $S_i$ measures the absolute integrated deformation activity during an above-threshold episode. Therefore, its distribution broadens at higher activity, giving a smaller fitted exponent $p_S$. This does not contradict the larger normalized kurtosis at low activity. It shows that at low activity, we have rarer, relatively more extreme deformation events. At high activity, we have more frequent and larger absolute above-threshold episodes.

\vskip 0.2cm
\noindent
\emph{Structure functions and multiscaling.} In order to characterize intermittency systematically, we look at the $p$th-order structure functions of the aspect-ratio increments, $S_p(\tau)=\big\langle |\Delta \mathrm{AR}(\tau;t)|^p\big\rangle$, for integer orders $1\le p\le 6$. In a scale-invariant regime, $S_p(\tau)\sim \tau^{\alpha_p}$. For self-similar Gaussian statistics, the $\alpha_p$ are linear in $p$, and any curvature in $\alpha_p$ signals intermittency~\cite{frisch1995turbulence,sreenivasan1997phenomenology}. Fig.~\ref{fig:structure_functions}(a) plots $\log S_p$ against $\log \tau$ at a representative activity. The curves are power laws only over an intermediate band of lags, from which we read off $\alpha_p$. At the shortest lags the signal is smooth, so $\Delta\mathrm{AR}(\tau;t)\approx\dot{\mathrm{AR}}(t)\,\tau$ and $S_p(\tau)\sim\tau^{p}$. At the longest lags they saturate once $\mathrm{AR}(t+\tau)$ and $\mathrm{AR}(t)$ decorrelate. This existence of a clean power-law band is the first indication of a nontrivial temporal scaling in the shape dynamics. 

The $\alpha_p$ grow monotonically with $p$, as seen in Fig.~\ref{fig:structure_functions}(b). However, we are interested in their departure from linearity. Therefore, we rescale by $\alpha_2$ and plot $\alpha_p/\alpha_2$. Similar analysis (extended self-similarity~\cite{benzi1993extended}) has been done in recent studies on multiscaling in active turbulence~\cite{benzi1993extended,mukherjee2023intermittency, kiran2025onset}. We expect that for a Gaussian self-similar process, $\alpha_p/\alpha_2=p/2$, as shown by the dashed line. However, all our data bends below this line at high orders, and the bend becomes increasingly pronounced as the activity is lowered. We quantify this further using the nonlinearity measure $\Delta_{\mathrm{NL}} = \sqrt{\frac{1}{N_p} \sum_{p=1}^{N_p} \left( \alpha_p - a\,p \right)^2}$. This provides the rms deviation of the exponents from the best linear fit $\alpha_p^{\mathrm{lin}}=ap$, with $a$ obtained from least-squares. $\Delta_{\mathrm{NL}}$ falls as $\zeta$ increases. Therefore, the deformation statistics drift towards linear, self-similar scaling at high activity while remaining strongly burst-dominated, in normalized units, at low activity.

\vskip 0.2cm
\noindent
\emph{Low-activity bath context.} It may appear counterintuitive that the stronger normalised deformation intermittency is at lower activity, since the mean deformation amplitude increases with activity. To clarify this point, we consider basic bath observables as functions of $\zeta$ (see the SM). As activity increases, both the bath velocity scale and defect density increase. However, the characteristic bath timescales, the turnover time $\tau_{\mathrm A}$, and the velocity-correlation time $\tau_v$ decrease with increasing $\zeta$. In other words, the active bath is slower but more temporally coherent. 

In line with this trend, we provide a physical interpretation of the deformation statistics discussed above. At low activity, the droplet is driven by forcing structures that are weaker in amplitude but longer lived. Therefore, larger deformation events are rarer but more coherent and more extreme \emph{relative to the typical fluctuation scale}. At higher activity, the bath is stronger but also fluctuates more rapidly. This increases the number of deformation events but reduces the intermittency of the normalized increments.

\vskip 0.2cm
\noindent
\emph{Role of active stresses and comparison with bath fluctuations.} The deformation bursts described above are driven by active forcing at the droplet interface. This raises the question of how directly large aspect-ratio increments track the interfacial active stress, and how the intermittency of $\mathrm{AR}(t)$ compares with that of the droplet's translation. Our observations suggest that instantaneous active stress is a poor predictor of instantaneous deformation. Conditioning the increment PDFs on large spatially averaged interfacial stress $\sigma_{\mathrm{act}}(t)$, the tails change weakly, enhancing them at low activity while suppressing them at high activity. Further, the correlation between $|\mathrm{d}\mathrm{AR}/\mathrm{d}t|$ and $\sigma_{\mathrm{act}}(t)$ is small everywhere (see SM). Lagged cross-correlations provide a sharper picture. The stress--speed correlation $C_{\sigma_{\mathrm{act}},U}$ peaks near zero lag, whereas the deformation-related correlations are weaker and broader in time (see SM). We therefore interpret that the translation is more directly slaved to bath forcing. Deformation bursts, on the other hand, reflect the joint action of the bath and the changing interface geometry. In other words, filtering places deformation below translation in a hierarchy of intermittency. Note that both these observables are non-Gaussian. However, the kurtosis ratio satisfies $\kappa_{\Delta\mathrm{AR}}/\kappa_{\Delta U} < 1$ across the range of activities and decreases with $\zeta$ (see SM). Therefore, translation increments have heavier tails, and the separation widens with increasing activity.

\begin{figure}[!ht]
	\centering
	\centering
	\includegraphics[width=1.0\linewidth]{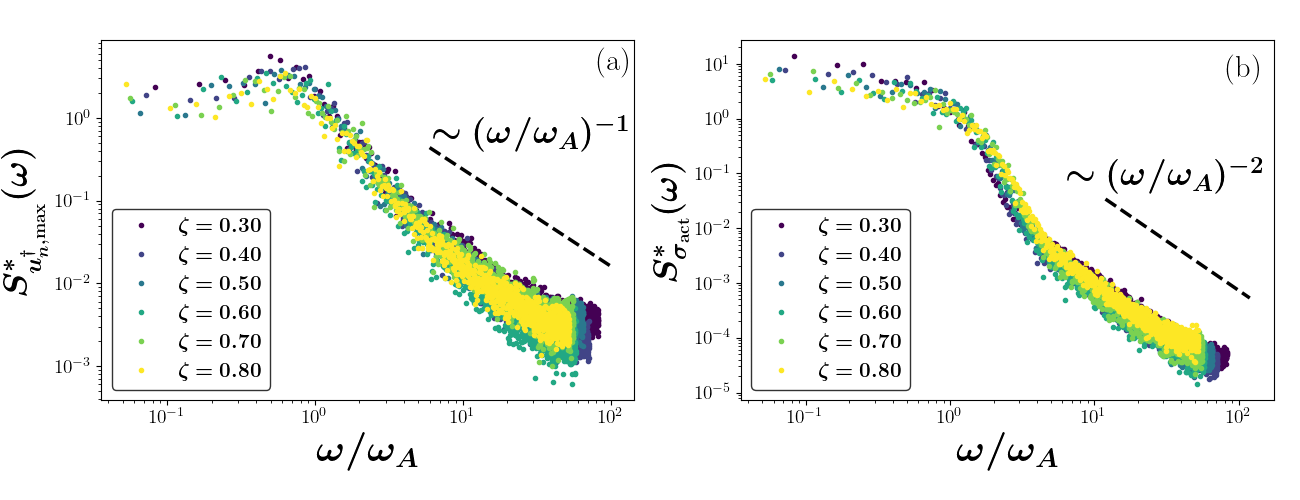}
    \caption{(Color online) Normalized power spectral densities versus the activity-scaled frequency $\omega/\omega_A$ for several activities at fixed interfacial tension $\gamma=7$. (a) Normal velocity $u_n^\dagger(t)$ at the distinguished interfacial point $s^\dagger(t)$ of maximal curvature; the dashed line is a guide to $\omega^{-1}$. (b) Spatially averaged interfacial active stress $\sigma_{\rm act}(t)$; the dashed line is a guide to $\omega^{-2}$.}
	\label{fig:spectral_signatures}
\end{figure}

\vskip 0.2cm
\noindent
\emph{Spectral signatures of deformation intermittency.} To probe intermittency further, we look at the power spectral densities (PSDs) of interfacial observables. For every trajectory, we identify the contour location $s^\dagger(t)$ of maximal curvature $\kappa(s,t)$ of the droplet and record the corresponding normal interface velocity $u_n^\dagger(t)$. In Fig.~\ref{fig:spectral_signatures}(a) we show the normalized PSD, $S^{*}_{u_n^{\dagger}{}_{,\max}}(\omega) = S_{u_n^{\dagger}{}_{,\max}}(\omega)/ \int S_{u_n^{\dagger}{}_{,\max}}(\omega)\, d\omega$ plotted against the rescaled frequency $\omega/\omega_{\mathrm A}$ for various activities. Over an intermediate range, the spectra show an approximate $\omega^{-1}$ scaling. The low-frequency part flattens due to finite observation time, and the high-frequency part is steeper due to microscopic relaxation. This $1/\omega$ behavior is observed in scale-free temporal correlations and intermittency~\cite{bak1988self,ditlevsen2010turbulence}. The normalization with the turnover timescale helps to partially align the spectra across activities, which suggests that this timescale is a useful organizing scale for the dominant interfacial fluctuations.. In Fig.~\ref{fig:spectral_signatures}(b), we similarly plot the normalized spectra of the spatially averaged interfacial active stress, $S^{*}_{\sigma_{\mathrm{act}}}(\omega) = S_{\sigma_{\mathrm{act}}}(\omega)/ \int S_{\sigma_{\mathrm{act}}}(\omega)\, d\omega$. This shows a steeper approximate $\omega^{-2}$ scaling. This corroborates that $1/\omega$ interfacial response does not directly reflect the active-stress signal, but is shaped by the interfacial tension. Additional quantities, such as localized interfacial stress spectra and simulations at different interfacial tensions (see SM), show that the response-side spectrum remains distinct from the forcing-side spectra and is modified by the mechanics of the interface.

The difference in the two spectra can further be phenomenologically addressed by comparison to a renewal-like burst process with heavy-tailed waiting times $\psi(\tau)\sim \tau^{-(1+\alpha)}$~\cite{lowen1993fractal}. Such processes are known to generate spectra $S^{*}(\omega)\sim \omega^{\alpha-1}$ or $S^{*}(\omega)\sim \omega^{-\alpha}$ depending on the signal type. In our study, the waiting time exponents satisfy $p_W\approx 0.7$--$1.3$ (Fig.~\ref{fig:AR_increments_bursts}(e)). Then, the predicted spectral slope is $\sim -1$, which is consistent with Fig.~\ref{fig:spectral_signatures}(a). This supports the suggestion that the droplet behaves like a nonlinear elastic filter, converting smoother forcing fluctuations into scale-free bursts of deformation. 

\vskip 0.2cm
\noindent 
\emph{Discussion.}
We have used fully resolved nematohydrodynamic simulations to study a passive droplet in extensile active turbulence. The droplet behaves as a deformable, activity-dependent probe whose shape fluctuations exhibit heavy-tailed increments, scale-free burst statistics, and multiscaling structure functions. Unlike droplets deformed by inertial turbulence, where breakup is governed by a balance of inertial stress and surface tension~\cite{hinze1955fundamentals, risso1998oscillations}, the present droplet is driven by defect-mediated active stresses at low Reynolds number. Although the mean deformation increases with activity, the normalized intermittency is strongest at low activity, consistent with slower and more coherent bath dynamics. 

Comparison with translational motion and forcing-side observables shows a hierarchy of intermittency: bath forcing is most intermittent, translation is less filtered, and deformation is the most buffered response due to interfacial tension and nematic elasticity. A hierarchy of timescales linked to intermittency has also been reported in other active systems~\cite{romaguera2025multiscale}. Correspondingly, the interface velocity exhibits a $1/\omega$ spectrum, distinct from the forcing-side $\omega^{-2}$ scaling, showing that the droplet response is not a trivial imprint of the active stress but emerges from nonlinear interface dynamics and burst-like temporal filtering.

Our results extend earlier studies of anomalous transport in active media~\cite{singh2024anomalous, padhan2023activity, kawakami2025migration},  and complement studies of activity-driven assembly of passive inclusions in active nematics~\cite{sariyar2026activity}, by showing that droplet \emph{shape} captures higher-order temporal statistics of the active bath. The measured multiscaling and burst exponents provide quantitative benchmarks for reduced models, which will likely require colored forcing, finite response times, and nonlinear bath--interface coupling beyond simple Gaussian noise descriptions.

Finally, our study is limited to two dimensions, a single droplet size, and a restricted parameter range. Since three-dimensional active nematics exhibit defect lines and loops rather than point defects~\cite{head2024spontaneous, digregorio2024coexistence,rorai2022coexistence}, we expect qualitatively new deformation and intermittency signatures in 3D. Exploring different interfacial and viscoelastic properties will also be important for understanding fluctuation filtering in soft active matter systems such as vesicles and biological condensates~\cite{reinken2025rheologically}.

\vskip 0.2cm
\noindent 
\emph{Acknowledgements.} The authors acknowledge the High Performance Computing Facility at IISER Mohali.


\vskip 1 cm



\appendix

\section{Model and parameters}
We model an active nematic bath containing a passive nematic droplet using incompressible active nematohydrodynamics for the velocity $\mathbf u(\mathbf x,t)$, pressure $p(\mathbf x,t)$, and symmetric traceless nematic tensor $\mathbf Q(\mathbf x,t)$:
\begin{equation}
\nabla\!\cdot\!\mathbf u = 0,
\end{equation}
\begin{equation}
\rho\left(\partial_t \mathbf{u} + \mathbf{u}\cdot\nabla \mathbf{u}\right)
= \nabla\cdot\boldsymbol{\sigma} - \mu \mathbf{u} + \mathbf{f}_I.
\label{eq:mom_end}
\end{equation}
Here $\mu$ is a substrate/friction coefficient and $\mathbf f_I$ is the interfacial force density. The total stress is
\begin{equation}
\boldsymbol{\sigma}= -p\mathbf{I} + 2\eta\mathbf{E} + \boldsymbol{\sigma}^{P} + \boldsymbol{\sigma}^{A},
\end{equation}
where $\mathbf{E}=(\nabla\mathbf{u}+(\nabla\mathbf{u})^{T})/2$ with active stress $\boldsymbol{\sigma}^{A} = -\zeta\,\mathbf{Q}$. The nematic evolves via the Beris--Edwards equation
\begin{equation}
\partial_t \mathbf{Q} + \mathbf{u}\cdot\nabla \mathbf{Q} - \mathbf{S} = \Gamma\,\mathbf{H},
\label{eq:Qeq_end}
\end{equation}
where $\Gamma$ is the rotational mobility, $\mathbf S$ is the standard co-rotation/stretching term, and $\mathbf H$ is the molecular field. In the single-elastic-constant approximation used here,
\begin{align}
\nonumber
\mathbf{H} = K\nabla^2\mathbf{Q} + \frac{C}{3}\mathbf{Q} + C\left(\mathbf{Q}\cdot\mathbf{Q}-\frac{\mathbf{I}}{d}\,\mathbf{Q}:\mathbf{Q}\right)\\ 
- C\,\mathbf{Q}\left(\mathbf{Q}:\mathbf{Q}\right),
\label{eq:Hfield_end}
\end{align}
where $K$ is the elastic constant and $C$ sets the bulk nematic energy scale. In the one-fluid formulation, material parameters are spatial fields that take distinct values in the two phases (active bath and passive droplet). In particular, $\zeta$ is nonzero in the outer active phase and set to zero inside the droplet, while density and viscosity take the phase-specific values listed in Table~\ref{tab:parameters}.

\section{Nondimensionalization and diagnostic groups}
We nondimensionalize using $x_0$ (length), $t_0$ (time), $u_0=x_0/t_0$ (velocity), and scale $\mathbf Q$ by a reference magnitude.
A convenient choice is
\begin{equation}
x_0=\sqrt{\frac{K_0}{\Gamma_0 C_0 \eta_0}},\qquad
t_0=\frac{1}{\Gamma_0 C_0},\qquad
u_0=\frac{x_0}{t_0},
\end{equation}
leading to the nondimensional groups
\begin{align}
\nonumber
\mathrm{Re}=\frac{\rho_0 u_0 x_0}{\eta_0},\qquad
\mathrm{Re}_a=\frac{t_0 \zeta_0}{\eta_0},\\
\mathrm{Re}_f=\frac{\mu_0 x_0^2}{\eta_0},\qquad
\mathrm{Re}_I=\frac{x_0 t_0 f_{I0}}{\eta_0}.
\label{eq:nd_groups_end}
\end{align}

To characterize the active bath, we use the root-mean-square velocity
\begin{equation}
u_{\rm rms}=\sqrt{\langle |\mathbf u|^2\rangle},
\end{equation}
and define an active turnover rate
\begin{equation}
\omega_A \equiv \frac{u_{\rm rms}}{\ell_c},
\end{equation}
where $\ell_c$ is a velocity-correlation length extracted from the fluid velocity autocorrelation. The corresponding turnover time is
\begin{equation}
\tau_A=\omega_A^{-1}.
\end{equation}
We also measure a bath velocity-correlation time $\tau_v$ from the temporal decay of the velocity autocorrelation. 

\section{Interfacial tension and front-tracking representation of $\mathbf f_I$}
We introduce capillarity as a force density concentrated on the interface,
\begin{equation}
\mathbf{f}_I(\mathbf{x},t)=\int ds\;\gamma\,\frac{d\mathbf{t}}{ds}\,
\delta\!\left[\mathbf{x}-\mathbf{x}_s(t)\right],
\label{eq:fI_cont_end}
\end{equation}
where $\gamma$ is the constant interfacial tension, $\mathbf t$ is the unit tangent, and $\mathbf x_s(t)$ parameterizes the interface. In the discrete front-tracking implementation, the Eulerian force applied to a cell of area $V=\Delta x\Delta y$ is approximated by summing tangent jumps at Lagrangian front points inside the cell~\cite{singh2024anomalous},
\begin{equation}
\frac{1}{V}\int_V dV\,\mathbf{f}_I \;\approx\;
\frac{\gamma}{\Delta x\Delta y}\sum_{i=1}^{N_f}\Delta\mathbf{t}_i.
\label{eq:fI_disc_end}
\end{equation}
This construction produces the Laplace pressure jump across the interface while retaining an Eulerian discretization for $\mathbf{u}$ and $\mathbf{Q}$.

\section{Numerical scheme and time stepping}
Eqs.~(\ref{eq:mom_end})--(\ref{eq:Qeq_end}) are integrated using a finite-volume discretization on a uniform Cartesian grid with a pressure-projection method which enforces incompressibility. Advection is computed using a high-order non-oscillatory reconstruction, and diffusive/elastic contributions use centered differences. The time step satisfies standard stability constraints associated with advection, viscosity, and capillarity:
 \begin{equation}
\Delta t = \min\!\left(
C_1\frac{\Delta}{u_{\textrm{max}}},\;
C_2\frac{2\eta_{\textrm{min}}}{\rho_{\textrm{max}}u_{\textrm{max}}^2},\;
C_3\frac{\rho_{\textrm{min}}\Delta^2}{4\eta_{\textrm{max}}}
\right),
\label{eq:dt_constraints_end}
\end{equation}
with $C_{1,2,3}=\mathcal O(1)$, $\Delta$ the grid size, and subscripts $\textrm{max, min}$ denoting the maximum and minimum values of the corresponding quantities across the two phases. The interface is represented by $N_f$ Lagrangian markers, and remeshed periodically to maintain nearly uniform arclength spacing.

\section{Numerical values}
The values of the various parameters are given in Table~\ref{tab:parameters}. The parameters are appropriately scaled using the definitions given earlier. 
\begin{table}[h]
\centering
\begin{tabular}{|c|c|}
\hline
\textbf{Non-dimensional Parameters} & \textbf{Values} \\
\hline
Domain size & $400 \times 400$ \\
\hline
Grid points & $200 \times 200$ \\
\hline
Droplet interfacial points, $N_f$ & $65$ \\
\hline
Density (active bath), $\rho_{\rm bath}$ & $0.10$ \\
\hline
Density (passive droplet), $\rho_{\rm drop}$ & $0.20$ \\
\hline
Viscosity (active bath), $\eta_{\rm bath}$ & $1.0$ \\
\hline
Viscosity (passive droplet), $\eta_{\rm drop}$ & $2.0$ \\
\hline
Radius of the droplet, $R$ & $12.0$ \\
\hline
Center of the droplet, $(x_c,y_c)$ & $(200.0,200.0)$ \\
\hline
Surface tension coefficient, $\gamma$ & $7.0$ \\
\hline
Flow-alignment parameter, $\lambda$  & $0.80$ \\
\hline
Activity (active bath), $\zeta_{\rm bath}$ & $0.10-0.80$  \\
\hline
Activity (passive droplet), $\zeta_{\rm drop}$ & $0.0$ \\
\hline
$\mu$  & $1.0$ \\
\hline
$\Gamma$  & $1.0$ \\
\hline
$C$  & $0.10$ \\
\hline
$K$  & $0.25$ \\
\hline
Time step,  $\Delta t$ & $0.001$ \\
\hline
$Re$ & $0.95$  \\
\hline
$Re_a$ & $1.0$ \\
\hline
$Re_f$ & $0.00075$ \\
\hline
$Re_I$ & $1.0$ \\
\hline
\end{tabular}
\caption{Non-dimensional simulation parameters for the active bath and passive droplet.}
\label{tab:parameters}
\end{table}

\section{Definitions used in the analysis}
Deformation is analyzed using aspect-ratio increments at lag $\tau$,
\begin{equation}
\Delta \mathrm{AR}(\tau;t)=\mathrm{AR}(t+\tau)-\mathrm{AR}(t),
\end{equation}
and the structure functions
\begin{equation}
S_p(\tau)=\langle|\Delta\mathrm{AR}(\tau;t)|^p\rangle_t.
\end{equation}
Averages are taken over stationary windows. In the main text, lag dependence may also be interpreted in terms of the normalized lag $\tau/\tau_A$.

For burst statistics, we use a base lag $\tau_b$ and a single threshold $\theta$ fixed across activities. This is chosen as a high quantile of $|\Delta \mathrm{AR}(\tau_b;t)|$ in a low-activity reference run. A burst is declared when $|\Delta \mathrm{AR}(\tau_b;t)|>\theta$, and consecutive above-threshold samples are merged into one episode. For episode $i$ we record
\begin{equation}
W_i=t^{\rm start}_{i+1}-t^{\rm end}_{i},
\qquad
S_i=\sum_{t\in i}|\Delta \mathrm{AR}(\tau_b;t)|,
\end{equation}
and report the CCDFs $P(W>w)$ and $P(S>s)$.

We define the scalar interfacial active-stress signal
\begin{equation}
\sigma_\text{act}(t) \equiv 
\Big\langle \big| \hat{\mathbf{n}}(s,t)\cdot \bm{\Pi}^\text{act}(\mathbf{r}(s,t),t)\cdot \hat{\mathbf{n}}(s,t) \big| \Big\rangle_{s\in \text{int}},
\end{equation}
with $\bm{\Pi}^{\rm act}=-\zeta\mathbf Q$. Conditional increment PDFs are formed by selecting times $t$ such that $\sigma_\text{act}(t)>\sigma^*$, where $\sigma^*$ is a fixed upper quantile of $\sigma_\text{act}$ in the stationary state. We then compute $\Delta \mathrm{AR}(\tau;t)$ on this subset (with $t+\tau$ restricted to the stationary time window).

To examine whether deformation is driven instantaneously or with a temporal lag, we also compute normalized lagged cross-correlations
\begin{equation}
C_{AB}(\tau)=
\frac{\langle \delta A(t)\,\delta B(t+\tau)\rangle}
{\sigma_A \sigma_B},
\end{equation}
with $\delta A=A-\langle A\rangle$, $\delta B=B-\langle B\rangle$, and $\sigma_A,\sigma_B$ the corresponding standard deviations. Positive $\tau$ therefore probes whether the response observable $B$ follows the forcing-side observable $A$ after a finite delay. We evaluate joint stress--deformation-rate statistics using
\begin{equation}
\dot{\mathrm{AR}}(t)\approx\frac{\mathrm{AR}(t+\Delta t)-\mathrm{AR}(t)}{\Delta t},
\end{equation}
and 2D histograms or binned means of $(\sigma_\text{act}(t),|\dot{\mathrm{AR}}(t)|)$.

For spectral analysis, we compare the power spectral density of the localized normal interface velocity $u_n^\dagger(t)$ with that of forcing-side observables. Here, the local interfacial point $s^\dagger(t)$ is chosen as the point of maximal curvature, along the contour. The corresponding localized forcing signal is
\begin{equation}
\sigma_{\rm act}^\dagger(t)=
\left|
\hat{\mathbf n}(s^\dagger,t)\cdot
\bm{\Pi}^{\rm act}(\mathbf r(s^\dagger,t),t)\cdot
\hat{\mathbf n}(s^\dagger,t)
\right|.
\end{equation}

\end{document}